\documentstyle[preprint,aps]{revtex}
\begin{document}
 \draft
 \title{\bf High-energy scissors mode}
\author{R. Nojarov,\thanks{
 Permanent address: Institute for Nuclear Research
 and Nuclear Energy, Bulgarian Aca\-de\-my of Sciences,
 BG-1784, Sofia, Bulgaria. Electronic address:
 nojarov@mailserv.zdv.uni-tuebingen.de}
 Amand Faessler\thanks{faessler@mailserv.zdv.uni-tuebingen.de}
 and M.~Dingfelder\thanks{dingfelder@mailserv.zdv.uni-tuebingen.de}
   }
\address{
Institut f\"ur Theoretische Physik, Universit\"at T\"ubingen, \\
Auf der Morgenstelle 14, D-72076 T\"ubingen, Germany}
\date{28 November 1994}
\maketitle
 \begin{abstract}
     All the orbital M1 excitations, at both low and high energies,
obtained from a rotationally invariant QRPA, represent the fragmented
scissors mode.  The high-energy M1 strength is almost purely orbital
and resides in the region of the isovector giant quadrupole
resonance. In heavy deformed nuclei the high-energy scissors mode is
strongly fragmented between 17 and 25 MeV  (with uncertainties
arising from the poor knowledge of the isovector potential). The
coherent scissors motion is hindered by the fragmentation and $B(M1)
< 0.25 \; \mu^2_N$ for single transitions in this region. The
$(e,e^{\prime})$ cross sections for excitations above 17 MeV are one
order of magnitude larger for E2 than for M1 excitations even at
backward angles.

 \end{abstract}
 \narrowtext
\section{Introduction}
   The nature of the orbital magnetic dipole (M1) strength in
deformed nuclei is still a controversial matter. The low-lying
orbital excitations with $K^{\pi} = 1^+$ have been interpreted
\cite{nofa88} as isovector rotational vibrations, i.~e. as a
manifestation of a weakly collective scissors mode \cite{nfscis}. A
strongly collective scissors state was predicted by the classical
picture of the Two-Rotor Model \cite{trm}, assuming that neutrons and
protons perform out-of-phase rotational oscillations around an axis
perpendicular to the nuclear symmetry axis. The fragmentation of the
M1 strength over a broad energy region, found experimentally between
2 and 4 MeV in heavy deformed nuclei (e.~g. Refs. \cite{pitz89}) and
present in microscopic theoretical calculations, hinders the
comparison with the simple classical picture.  It has been shown
earlier \cite{nofa88,nofa94} that the energy and the amount of the
high-lying M1 strength, corresponding to $\Delta {\cal N}$ = 2
excitations, are strongly dependent on the type of the residual
quadrupole-quadrupole (or non-spin) interaction used in microscopic
calculations within the  quasiparticle random-phase approximation
(QRPA).

    The above interpretation in terms of scissors mode is supported
by some theoretical works \cite{trmoya,iudice,otsuka,heyde91,ikeda}
and questioned by others \cite{hama,magnus,speth,zaw90,zaw93}.  The
scissors mode is found in Refs. \cite{speth,zaw90,zaw93} at high
energy, E$_{\rm theor} \approx 22-24$ MeV, as a part of the isovector
giant quadrupole resonance (IVGQR).  These three works use
Landau-Migdal interactions in QRPA.  Similar conclusions are obtained
in QRPA with separable forces \cite{hana}, where the IVGQR lies at
E$_{\rm theor} \approx 27$ MeV and overlaps 40\% with the scissors
mode. It was shown \cite{zaw90} that the collective hydrodynamical
model predicts a high-energy scissor mode, but the low-lying state is
spurious. However, after taking additionally the nuclear elasticity
within the same classical model into account, a low-energy mode was
obtained as well \cite{zaw93}. The isovector rotor model \cite{irm}
predicts the appearance of the scissors mode at both low and high
energies.

   We are going to discuss in Sec. III the isovector coupling of this
interaction, which influences strongly the high-lying orbital M1
strength. An isoscalar symmetry restoring interaction is introduced
in Sec. II, together with an isovector interaction.  The E2 strength
distribution is studied in Sec. III in relation to the isovector
coupling constant. The high-energy M1 strength and related
$(e,e^{\prime})$ cross sections are discussed in Sec.  IV. The
conclusions are summarized in Sec. V.

\section{Decoupled isovector interaction}

   The deformed mean field ${\bf H}_0$ is obtained in our formalism
\cite{nofa88,fano90} by diagonalizing an axially-symmetric
Woods-Saxon potential. Electric and magnetic transition operators are
treated on equal footing by introducing a signature $m = \pm 1$
\cite{nofa88}, corresponding to the directions $x$ and $y$,
indistinguishable in the intrinsic frame. For instance, the total
angular momentum operator, ${\bf J}_t \equiv {\bf J}(m,t)$, defined
by Eq. (4) of \cite{nofa94}, has quasiparticle (q.p.) matrix elements
$j^t_{ki} \equiv j(ki,mt)$ \cite{nofa93}.  The isospin index $t$
denotes neutrons or protons. We shorten the notations in the
following by omitting, wherever possible, the signature index $m$,
which is of technical interest only.

   Following the procedure of Pyatov \cite{baznat}, we construct in
the quasiboson approximation operators of quadrupole type ${\bf F}$.
They are used to define the isoscalar interaction ${\bf H}_{S}$
\cite{nofa88,nofa94}, which restores the rotational symmetry of the mean
field ${\bf H}_0$, violated by the deformation:
\begin{eqnarray}
{\bf F}_t \equiv {\bf F}(m,t) = [ {\bf H}_0, {\bf J}_t ], \nonumber
\\  {\bf F}_S = {\bf F}_n + {\bf F}_p, \quad {\bf F}_V = {\bf F}_n -
{\bf F}_p, \nonumber \\
{\bf H}_S = - { k_S \over 2} \sum_m m {\bf F}_S^2(m), \label{his}
\end{eqnarray}
Under the assumption for a simple quadrupole deformation ${\bf
Q}_{20}$ only, as in the case of the Nilsson potential, the operators
${\bf F}_S(m)$ (\ref{his}) are proportional to the quadrupole
operators ${\bf Q}_{2,\pm 1}$.

  A separable residual interaction of the general form,
\begin{eqnarray}
{\bf H}_{FF} = - {\textstyle{ 1 \over 2 }} \sum_m m \Bigl [ k_{nn}
{\bf F}^2_n(m)  +  k_{pp} {\bf F}^2_p(m)  \nonumber \\
+  2k_{np} {\bf F}_n(m) {\bf F}_p(m) \Bigr ], \label{hff}
\end{eqnarray}
can be decomposed into a sum of isoscalar, isovector, and coupling
terms with the corresponding coupling constants,
\begin{eqnarray}
k_0 = {\textstyle{ 1 \over 4 }} (k_{nn} + k_{pp} + 2 k_{np} ),
\nonumber \\
k_1 = {\textstyle{ 1 \over 4 }} (k_{nn} + k_{pp} - 2 k_{np} ),
\nonumber \\
k^{\prime} = {\textstyle{ 1 \over 2 }} (k_{nn} - k_{pp}), \quad
r^{\prime} = k_1/ k_0. \hfill \label{svint}
\end{eqnarray}
The isovector operator ${\bf F}_V$ (\ref{his}) is obtained from the
isovector angular momentum ${\bf J}_n - {\bf J}_p$. However, it is
more convenient to construct a relative isovector interaction ${\bf
H}_R$ and the coupling ${\bf H}_C$ with the isoscalar term
(\ref{his}) by introducing a relative angular momentum ${\bf J}_R$:
\begin{eqnarray}
{\bf J}_R = \sqrt{ X_p \over X_n} {\bf J}_n -  \sqrt{ X_n \over X_p}
{\bf J}_p, \quad
X_t = \langle [ {\bf J}_t^{\dag}, {\bf F}_t ] \rangle, \nonumber \\
 {\bf F}_R \equiv {\bf F}_R(m) = [ {\bf H}_0, {\bf J}_R (m)],
 \nonumber \\
{\bf H}_R = - { k_R \over 2} \sum_m  m {\bf F}_R^2(m), \nonumber \\
{\bf H}_C = - { k_C \over 2} \sum_m  m {\bf F}_R(m) {\bf F}_S(m).
\label{hr} \end{eqnarray}
It is easy to show that the relative isovector interaction ${\bf
H}_R$ is rotationally invariant and its coupling constant $k_R$
remains a free parameter. This interaction is slightly more symmetric
than that used in Ref. \cite{magnus}, but the difference is
insignificant.  The condition for rotational symmetry provides the
same value for the isoscalar constant $k_S$ as in Eq. (9) of
\cite{nofa94} and closes the coupling ${\bf H}_C$ between the
isoscalar and relative channels:
\begin{eqnarray}
[ {\bf H}_0 + {\bf H}_S + {\bf H}_R + {\bf H}_C, \; {\bf J} ] = 0,
\nonumber \\
\Longrightarrow k_S = 1/X, \; k_C =0, \quad X = X_n + X_p.
\label{inv}
\end{eqnarray}
Thus, the use of relative interaction enables the decoupling of the
two channels and the symmetry-restoring procedure \cite{nofa88,nofa94}
is no more necessary, since the hamiltonian (\ref{inv}) is already
rotationally invariant.

  We introduce further a spin-spin residual interaction ${\bf
H}_{SS}$ \cite{fano90}, which is always rotationally invariant and
does not change the coupling constants (\ref{inv}). The latter can be
related to the coupling strengths of the quadrupole interaction in
the general form (\ref{hff}). Thus, our q.p. model hamiltonian, used
in the present calculations, has the following form:
\begin{eqnarray}
 {\bf H} = {\bf H}_0 + {\bf H}_{FF} + {\bf H}_{SS},
\nonumber \\
 k_{nn} = k_S (1 + r {X_p \over X_n}), \quad
  k_{pp} = k_S (1 + r {X_n \over X_p}), \nonumber \\
  k_{np} = k_S (1 - r), \quad r = { k_R \over k_S}, \label{ham}
\end{eqnarray}
where $k_S$ is calculated microscopically (\ref{inv}) and the ratio
$r$ is treated as a free parameter to be discussed in the next
section. The relationships with the constants $k_0, \; k_1$, and
$k^{\prime}$ are obtained upon insertion of (\ref{ham}) into
(\ref{svint}). The constants of the spin-spin interaction,
\begin{equation}
c(+)A = 200 \; \hbox{MeV}, \quad c(-) = -0.5 c(+), \label{sc}
\end{equation}
are derived from nuclear matter calculations in the way described in
Ref. \cite{noj93}.  The deformation-independent Woods-Saxon
parameters are taken from Ref. \cite{tanaka}.  When applied to other
rare-earth \cite{sarri} and actinide \cite{noj93} nuclei, this
parametrization has lead to a good agreement with experimental data
for single M1 transitions at low energy, observed in ($e,e^{\prime}$)
and ($\gamma ,\gamma^{\prime}$) experiments.  The energy
distribution of the spin-flip M1 strength between 6 and 10 MeV,
determined by inelastic proton scattering, has also been reproduced
in \cite{noj93,sarri}.

   The QRPA equations of motion are solved for $K^{\pi} =1^+$
excitations in the intrinsic frame using the model hamiltonian
(\ref{ham}).  We present here as an example the results for
$^{160}$Gd, obtained with deformations $\beta_2 = 0.26, \; \beta_4 =
0.025$ and pairing gaps $\Delta_n = 0.8 \; {\rm MeV}, \; \Delta_p =
1.1 \; {\rm MeV}$. Expressions for M1 and E2 transition probabilities
in terms of RPA amplitudes are given in Ref. \cite{nofa88}. The B(M1)
values  for transitions to $I^{\pi}K =1^+1$ states are calculated
with bare orbital and effective spin gyromagnetic factors, $g_s = 0.7
g_s^{\rm free}$. The B(E2) values  for transitions to the
corresponding $I^{\pi}K =2^+1$ states are obtained without
introducing effective charges, i.~e. they are of purely proton
nature.

   Our attention shall be focused in the next section on the ratio
$r$ (\ref{ham}), determining the strength of the isovector
spin-independent interaction.

\section{Isovector coupling constant}

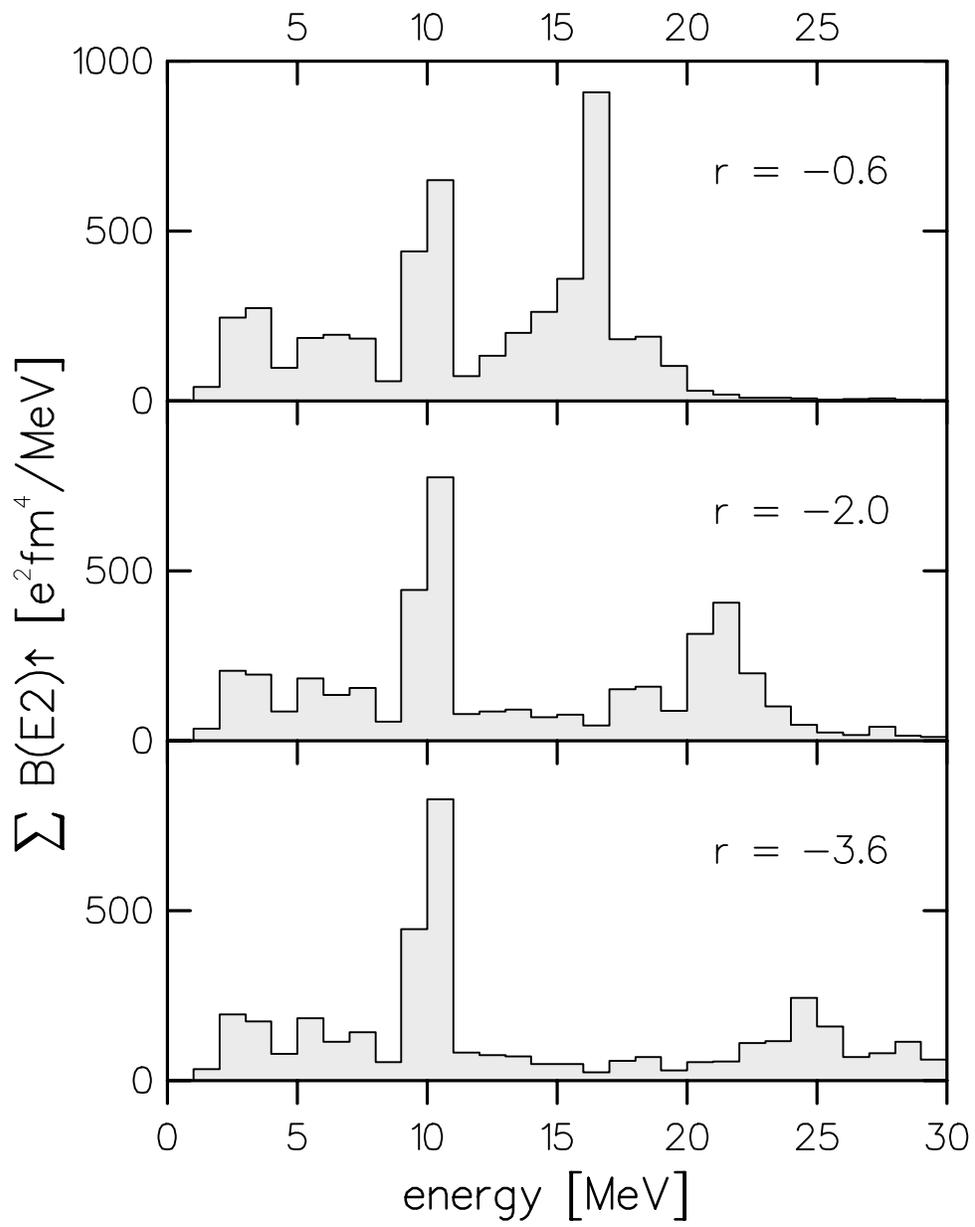
\begin{figure}
\caption{Energy distribution of $B(E2;0^+_{\rm g.s.} \to 2^+1)$ values,
summed in bins of 1 MeV, for $^{160}$Gd, corresponding to three
different values of the isovector strength constant $k_R$
(\protect\ref{ham}): $r = -0.6, \; -2.0, \; -3.6$. }
\label{fig1} \end{figure}

   The energy distribution of the E2 strength with $K=1$ is shown in
Fig.~1 as histograms for three different values of the isovector
coupling constant $k_R$ (\ref{ham}). The latter is fixed by the ratio
$r = k_R/k_S$, because the isoscalar constant $k_S$ is determined
microscopically from (\ref{inv}). The lower bump has a peak at about
11 MeV, which is smeared out by the histogram. It represents the
$K=1$ component of the isoscalar giant quadrupole resonance (ISGQR),
observed by inelastic electron scattering \cite{woude}. Its position,
determined by the isoscalar constant $k_S$, agrees with the
prediction of the vibrating potential model \cite{suz77},
\begin{equation}
E(21;is) \approx {63 \over A^{1/3} } (1- {\delta \over 6} ) \;
\hbox{MeV}. \label{ise}
\end{equation}
The summed contribution of the $K = \pm 1$ components to the
classical isoscalar and isovector energy-weighted (e.w.) sums are
\cite{warb}
\begin{eqnarray}
S(E21;is)_C = S(E21)_C Z/A, \nonumber \\
S(E21;iv)_C = S(E21)_C N/A, \nonumber \\
S(E21;is)_C + S(E21;iv)_C = S(E21)_C, \nonumber \\
S(E21)_C = { 5 e^2 \hbar ^2 \over 2 \pi m} Z \langle r^2 \rangle _p.
\label{cle2} \end{eqnarray}
We calculate microscopically the mean-square-radius, $\langle r^2
\rangle _p = 27.3 \; fm^2$, and obtain with this value from
(\ref{cle2}) the classical e.w. sum $S(E21)_C = 57573 \; e^2fm^4$MeV.
The RPA isoscalar E2 e.w. strength between 9 and 11 MeV, shown in the
middle plot of Fig.\ \ref{fig1}, is $S(E21; 9-11 \; \hbox{MeV})_{\rm
th} = 12215 \; e^2fm^4$MeV, i.e.  only 53\% from the classical
isoscalar e.w. sum $S(E21;is)_C = 23029 \; e^2fm^4$MeV are exhausted.
This result corroborates with the experimental data for other heavy
deformed nuclei \cite{woude}.  However, no strength is missing, since
the total RPA e.w. strength, $S(E21; 0-30 \; \hbox{MeV})_{\rm th} =
57272 \; e^2fm^4$MeV, exhausts 99.5\% of the above classical value
$S(E21)_C$.  One should note further that the decomposition
(\ref{cle2}) of the classical E2 strength into isoscalar and
isovector parts is a rather crude approximation, because it does not
take into account the redistribution of these two components caused
by the residual interactions.

   It is seen from Fig.\ \ref{fig1} that the low energy E2 strength
distribution and the position of the ISGQR are rather insensitive to
the isovector residual interaction. The latter is repulsive and
concentrates the isovector strength in the higher-lying bump,
consisting of $\Delta {\cal N} = 2$ excitations. The non e.w. E2
strength at high energy decreases for a stronger isovector
interaction, because the e.w. E2 strength is conserved in the three
plots of Fig.\ \ref{fig1}. This is due to the fact that the
interaction (\ref{hff}) redistributes the E2 strength, but does not
contribute to the E2 e.w. sum rule. The latter is determined by the
mean field and the spin-spin interaction \cite{fano90}, which are
fixed in the three cases of Fig.\ \ref{fig1}.

   The isovector coupling constant $k_R$ could be determined from the
position of the second maximum, representing the isovector giant
quadrupole resonance (IVGQR). Experimental data on it for heavy
deformed nuclei are scarce \cite{expiv}. The measured
$(e,e^{\prime})$ spectrum is usually fitted by a number of gaussians.
That one, lying closest to the  theoretically expected energy
\cite{suz77},
\begin{equation}
E(21;iv) \approx {141 \over A^{1/3} } (1- {\delta \over 6} ) \;
\hbox{MeV}, \label{ive}
\end{equation}
is identified with the IVGQR. The prediction (\ref{ive}) results from
evaluation of the isovector coupling constant within the collective
model. The main uncertainty of this procedure originates from the
isovector potential $V_1$, which is estimated by relating the
isovector density with the potential asymmetry energy
\begin{eqnarray}
{ V_1 \over 8 } \int { (\rho_n - \rho_p)^2 \over \rho } {\rm d} {\bf r}
 = K_{as} { (N-Z)^2 \over A }, \label{int} \\
V_1 = 8 K_{as}. \label{ivpot}
\end{eqnarray}
Using a recent empirical value, $K_V = 30.6$ MeV \cite{nix}, for the
volume asymmetry energy from the Bethe-Weizs\"acker formula, and
subtracting the kinetic energy  $b_{kin} = 2 K_{kin} = 25$ MeV
\cite{brm}, one obtains $K_{as} = K_V - K_{kin} \approx 18$ MeV. The
resulting isovector potential  $V_1 = 144$ MeV (\ref{ivpot}) is a
little bit larger than the previous estimate of 130 MeV \cite{brm},
based on a smaller value of $K_V$.

  The identity (\ref{ivpot}) follows from (\ref{int}) under the
assumption that the integral in the l.h.s. of (\ref{int}) is equal to
$(N-Z)^2 / A$. This is true for constant neutron and proton densities
within a sphere with a sharp surface. When the integral is calculated
with more realistic Woods-Saxon densities and the isovector potential
is determined from (\ref{int}), instead of (\ref{ivpot}), we obtain
 $V_1 = 67.5$ MeV for $^{160}$Gd. Test calculations for several more
rare-earth nuclei have lead to values close to 68 MeV.

  The ratio between the isoscalar and isovector coupling constants of
the quadrupole interaction can be calculated within the collective
model \cite{suz77}.  Using our microscopic mean square radius for
$^{160}$Gd, the above isovector potential  $V_1 = 67.5$ MeV, and an
oscillator constant $\hbar \omega_0 A = 41$ MeV, we obtain for this
ratio $r^ {\prime} = k_1/k_0 = -1.5$. However, as discussed in
Sec.~II, the operators ${\bf F}$ (\ref{his}) do not coincide with the
quadrupole operators. Moreover, the isoscalar coupling constant $k_S$
is already determined from the rotational invariance (\ref{ham}). The
isovector constant $k_R$ can be evaluated in our case following the
procedure of Ref. \cite{suz77} for the operator ${\bf F}$:
\begin{eqnarray}
k_R = - {V_1 \over 8} \left [ \int {\bf F}^{\dag} \rho {\bf F}
d{\bf r} \right ]^{-1} = - {V_1 \over 8} \langle {\bf F}{\bf F}^{\dag}
 \rangle ^{-1}, \nonumber \\
\langle {\bf F}{\bf F}^{\dag} \rangle = 2 \sum_{ik} E^2_{ik}
j^2_{ik}, \label{kr}
\end{eqnarray}
where the summation runs over neutrons and protons. An additional
factor of 1/2 is present in (\ref{kr}) in comparison  with
\cite{suz77}, since our coupling constants (\ref{hff},\ref{svint})
are two times smaller than usual. This is due to the
$m$-symmetrization of the interaction ${\bf H}_{FF}$
(\ref{hff},\ref{ham}). Using $V_1 = 67.5$ MeV, one obtains from
(\ref{inv},\ref{kr}) the ratio $r = k_R / k_S = - 1.67$, consistent
with the above value of $r^{\prime}$ for a quadrupole interaction.

   These numbers contain still rather large uncertainties.
Corrections to the isovector potential could arise, e.~g. from the
surface asymmetry energy. It contributes even to the IVGQR
\cite{suz80}, regarded as a predominantly volume oscillation. Further
renormalization effects \cite{bes} can lead even to attractive
n-n and p-p quadrupole interactions with $r^{\prime} = -0.6$. Thus,
the theoretical estimates of the isovector coupling constant are not
accurate enough and only the experimental data on the IVGQR could
decide on this issue. We adopt the value $r = -2$, producing with our
potential an IVGQR centered at about 22--23 MeV, as seen from the
middle plot of Fig.\ \ref{fig1}.

This result agrees with experimental findings in several rare-earth
nuclei \cite{expiv} and more reliable data on $^{208}$Pb
\cite{woude}.  The theoretical E2 strength between 17 and 25 MeV,
shown in the same plot and corresponding to the IVGQR with $K^{\pi} =
1^+$, amounts to 1475 $e^2 \; fm^4$.  The value is close to a rough
estimate \cite{iudice} from a schematic two-level model and agrees
qualitatively with recent results on the IVGQR in $^{154}$Sm,
obtained from inelastic scattering of polarized protons \cite{achim}.
The microscopic isoscalar coupling constant $k_S$ (\ref{inv}) gives
rise to the isoscalar GQR with $K^{\pi} = 1^+$ at 11 MeV. Its E2
strength from the middle plot of Fig.\ \ref{fig1}, summed between 9
and 11 MeV, is $B(E2) = 1193 \; e^2fm^4$, in agreement with
experiment \cite{woude,expiv}.

\section{High-energy M1 strength and ($\lowercase{e},\lowercase{e}
 ^{\prime})$ cross sections}

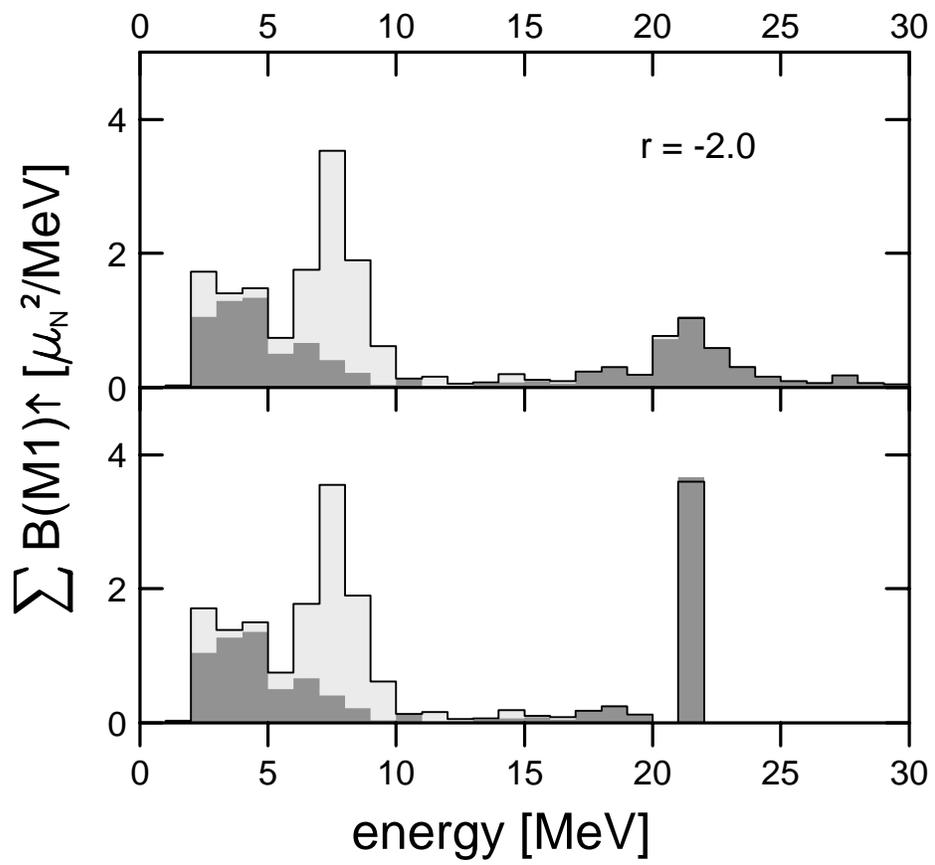
\begin{figure}
\caption{Energy distribution of $B(M1;0^+_{\rm g.s.} \to 1^+1)$
values summed in bins of 1 MeV (contour histograms with a bright shading)
for $^{160}$Gd, obtained with isovector strength ratio (\protect\ref{ham})
$r = -2$.  Dark-shaded areas: only orbital M1 strength. Bottom plot: only
2q.p. configurations below 20 MeV are included in calculations.}
\label{fig2} \end{figure}

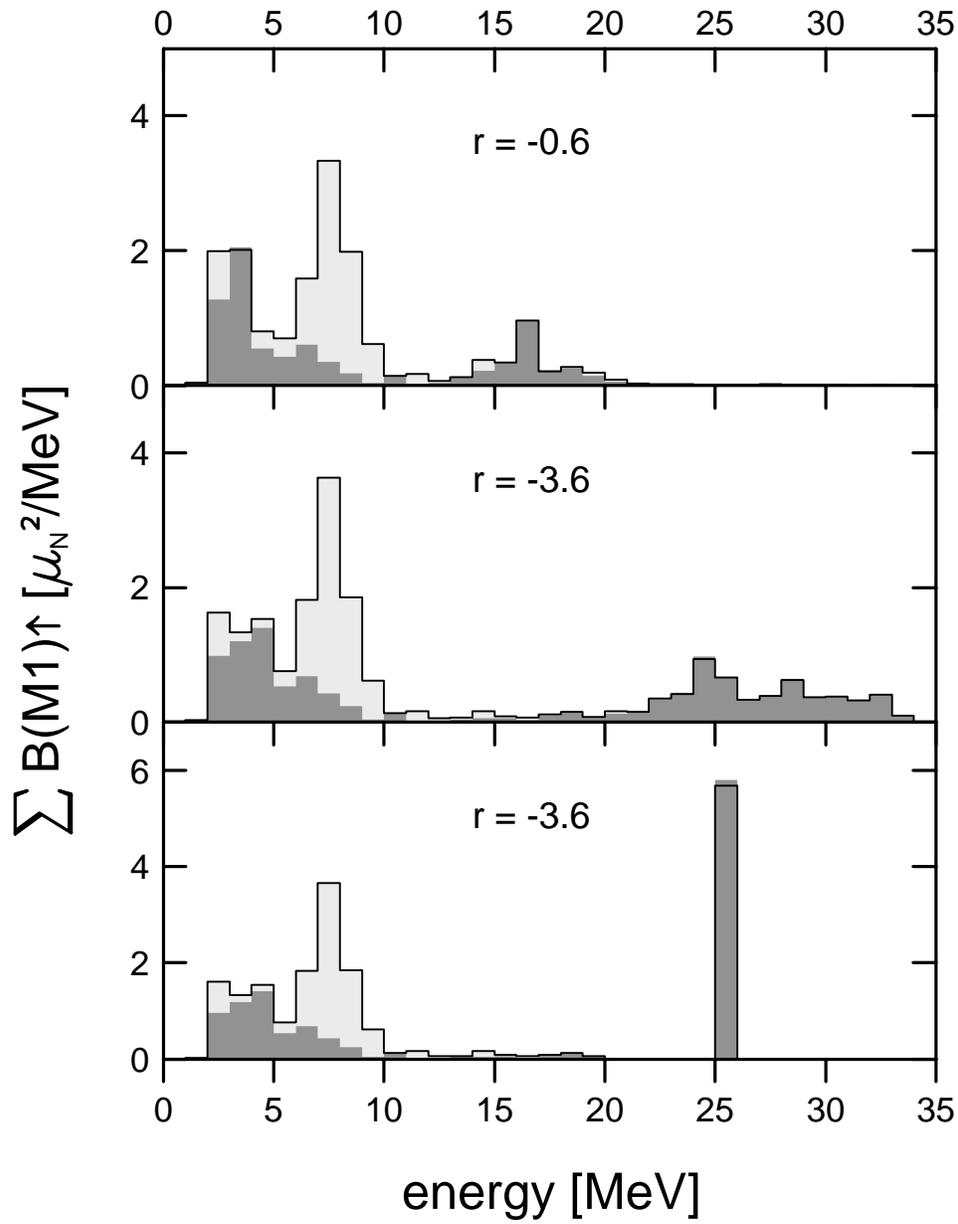
\begin{figure}
\caption{The same as in Fig.\ \protect\ref{fig2}, but for isovector
strength ratios $r = -0.6, \; -3.6$. Basis cut-off in the bottom plot
as in the lower plot of Fig.\ \protect\ref{fig2}. Note the compressed
ordinate scale in the bottom plot.}
\label{fig3}  \end{figure}

   The energy distribution of M1 strength in $^{160}$Gd, summed
in bins of 1 MeV, is shown as histograms in Figs.\ \ref{fig2} and
\ref{fig3} for different isovector ratios $r$ (\ref{ham}). The total
and orbital M1 strengths  are represented by contour lines and shaded
areas, respectively.  It is seen that the high-energy M1 strength is
purely orbital, while the low-energy one contains spin contributions,
which are dominant between 6 and 10 MeV.  Comparison with the
corresponding cases from Fig.\ \ref{fig1} shows that the high-lying
orbital M1 strength resides in the same energy region as the E2
IVGQR.

   The low-lying orbital M1 strength is influenced only slightly when
increasing the magnitude of the isovector interaction from $r=0$ to
$r \approx -1$. A further increase  ($r < -1$) does not produce
appreciable changes.  The quadrupole interaction (\ref{hff})
contributes to the M1 e.w.  sum-rule \cite{nofa88}. The repulsive
isovector interaction shifts the high-lying orbital strength to
higher energy and increases the e.w. sum.

   Figure \ref{fig2} displays results corresponding to the adopted
isovector coupling constant, $r=-2$.  The lower plot of Fig.\
\ref{fig2} was obtained by including only 2q.p.  configurations below
20 MeV  in the RPA calculations, as done also in \cite{hana}. A
schematic, single excitation is produced at 22 MeV, carrying most of
the high-energy strength: $B(M1) = 3.6 \; \mu^2_N, \ \; B(E2) = 1305
\; e^2fm^4$.  However, this strongly collective state is no more
present in more realistic calculations, including further 2q.p.
configurations up to 30 MeV,  as seen from the upper plot of Fig.\
\ref{fig2}. The single excitation is strongly fragmented over a
broad energy region in this case.

   Comparison with our previous results from Figs. 3 and 4 of Ref.
\cite{nofa94} demonstrates that the strong fragmentation is not
solely due to the large basis, but results also to a great extent
from the restoration of rotational invariance. It is seen from these
two figures that the quadrupole interaction alone produces a
high-energy orbital M1 strength, which is still strongly concentrated
in a narrow peak. In contrast, considerable fragmentation takes place
after adding the symmetry-restoring interaction \cite{nofa88} (cases
denoted by a prime in Figs. 3,4 of Ref. \cite{nofa94}). This strong
fragmentation is in a qualitative agreement with the present results,
obtained from the rotationally invariant hamiltonian (\ref{ham}).

   The summed M1 strength above 20 MeV in the upper plot of Fig.\
\ref{fig2} is equal to that of the single excitation at 22 MeV from
the bottom plot, resulting from the basis cut-off.  The equality is
due to the fact that the high-energy strength originates from 2q.p.
configurations with $\Delta {\cal N} =2$, lying at about 15 MeV.
Thus, configurations above 20 MeV do not generate but only
redistribute the high-energy strength, pushed by the repulsive
isovector interaction from 15 to 22 MeV.

  Let us consider the more realistic results in the upper plot of
Fig.\ \ref{fig2}.  The summed (predominantly orbital) M1 strength
between 17 and 25 MeV is 3.6 $\mu^2_N$, in agreement with a recent
estimate for $^{154}$Sm from $(\vec{p}, \vec{p}\; ^{\prime})$
experiments \cite{achim}. It is derived from a sum-rule \cite{moza}
connecting the M1 and E2 strengths.  The strongest single 1$^+$
excitation from the considered energy region lies at 22.5 MeV and
carries only 0.25 $\mu^2_N$ strength, i.~e. it could be hardly
resolved experimentally.  This excitation overlaps only 4\% with the
microscopic scissors state \cite{trmoya,nfscis}, obtained by rotating
the neutron and proton configurations of the deformed mean field in
opposite directions. The low-lying 1$^+$ excitations below 9 MeV
overlap all together 77\% with the scissors state, while the overlap
of those between 17 and 25 MeV amounts to 55\%. The total overlap of
all the RPA excitations below 30 MeV is 1.46.  Thus, the low- and
high-lying excitations exhaust 53\% and 38\% of the total overlap
with the scissors state, respectively.

    Let us note that the total overlap is exactly 100\% for a
vanishig isovector interaction \cite{nfscis,nofa94}.  The isoscalar
channel cancels the spurious (isoscalar) admixtures, but it does not
contribute to the (purely isovector) scissors mode.  Thus, an overlap
of 100\% is produced by the mean field alone. The overlap is not
conserved in RPA since it corresponds to a non e.w. sum, while only
energy-weighted sums are conserved in RPA. The excess beyond 100\% is
generated by the isovector residual interaction, which is not present
in the microscopic scissors state, constructed from the mean field
alone.  The overlap reaches 174\% in the case $r = -3.6$, plotted in
Fig.\ \ref{fig3}.

   A more meaningful geometric interpretation of the orbital 1$^+$
excitations is provided by the comparison with low collective
synthetic states \cite{fanos}. A synthetic state is constructed for
each RPA excitation by rotating out-of-phase neutron and proton
configurations weighted with the RPA amplitudes of this excitation.
A large overlap of a given RPA state with its synthetic
counterpart is an indication for an isovector rotational motion,
irrespectively of the low collectivity caused by the fragmentation.
The strongest (low-energy orbital) M1 excitation overlaps usually
80--90\% with its synthetic counterpart \cite{nfscis,noj93}. In
contrast, no one of the strongest orbital states at high energy
(around 22 MeV in the upper plot of Fig.\ \ref{fig2}) overlaps more
than 7\% with its synthetic counterpart. The coherence is restored in
the schematic single excitation at 22 MeV in the lower plot: it
overlaps almost 90\% with its synthetic counterpart, confirming the
presence of the scissors mode at high-energy.

   It is clear from the above considerations that the scissors mode
fragments over both the low- and high-energy orbital M1 excitations.
This holds exactly for the isovector rotational model \cite{irm},
formulated  in the schematic two-level basis of the quantized
deformed oscillator.  It was found within this model that the
scissors mode exhausts the whole non-spurious orbital M1 strength at
low and high energy, corresponding to excitations with $\Delta {\cal
N}$ = 0,2, respectively.

   A strong isovector quadrupole interaction, based on the operators
${\bf Q}_{2,\pm 1}$, leads to similar results  \cite{nofa94}, apart
from  concentrating in a narrower energy region much more high-lying
M1 strength. Most of it turns out to be spurious. After restoration
of the rotational invariance, the high-lying strength is reduced and
fragmented over a broad region \cite{nofa88,nofa94}, in agreement
with the present results, based on the rotationally invariant
hamiltonian (\ref{ham}).

   Results from different QRPA calculations are compared in Table~I.
Our results (first row) are similar to those obtained with a zero-range
Landau-Migdal interaction \cite{speth,zaw90} (second row), apart from
a lesser fragmentation in the latter case.  This is seen on the
example of $^{164}$Dy \cite{zaw90}, where the strongest high-lying
1$^+$ state (at 21 MeV) has a large transition probability, $B(M1) =
1.26 \; \mu^2_N$. It seems that the concentration of strength is
caused by the reduced basis of 2q.p. configurations adopted in Ref.
\cite{zaw90}. As discussed above, the total strength is not
influenced substantially by this procedure.

   The results listed in the last row of Table~I \cite{hana} are
similar to those in the bottom plot of Fig.\ \ref{fig3}, because the
same basis cut-off (20 MeV) and almost the same isovector ratio are
used in both cases.  The B(M1) value of the single excitation at 26
MeV in the bottom plot of Fig.\ \ref{fig3} is four times larger than
the value 1.5 $\mu_N^2$, listed in the last row of Table~I. The
difference is due to the fact that $^{152}$Dy, considered in
\cite{hana}, has a very small deformation, producing less orbital M1
strength than the well-deformed $^{160}$Gd, studied in the present
work.

   It is seen from the middle plot of Fig.\ \ref{fig3} that a
considerable fragmentation takes place after including  all the 2q.p.
configurations up to 35 MeV in calculations. The strong collective
excitation disappears and its strength is distributed over a wide
energy range.  As discussed in Sect.~III, the present status of
experiment on heavy deformed nuclei does not allow a reliable
discrimination between the results in Figs.\ \ref{fig2} and
\ref{fig3}, obtained with different isovector coupling constants.

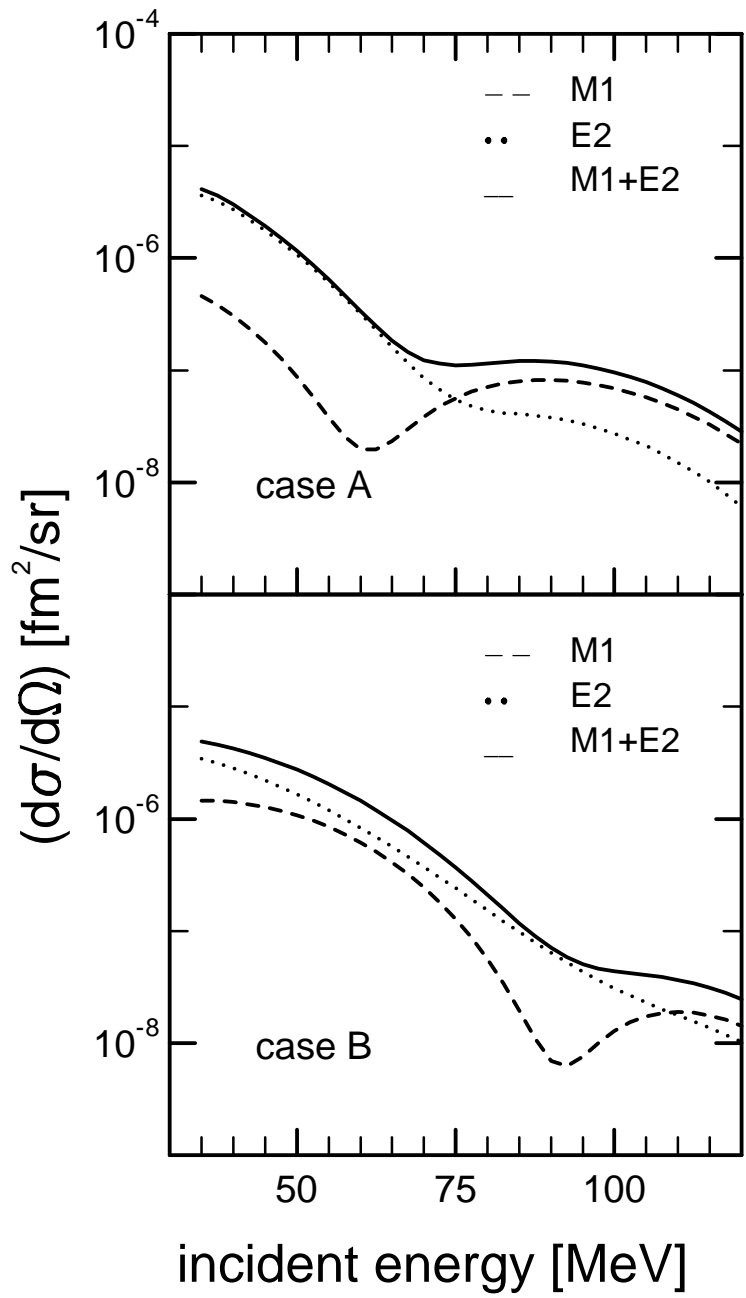
\begin{figure}
\caption{DWBA $(e,e^{\prime})$ cross sections of the strongest (case
A) and the second strongest (case B) RPA excitations around 22 MeV
from the upper plot of Fig.\ \protect\ref{fig2} with $B(M1)$ values
0.25 and 0.20 $\mu^2_N$, respectively. Scattering angle $\theta =
165^{\circ}$. Contributions from E2$(0^+_{\rm g.s.} \to 2^+1)$ (dotted
curves) and M1$(0^+_{\rm g.s.} \to 1^+1)$ (dashed curves) excitations
of the same rotational band with $K^{\pi}=1^+$, and  their sum
(continuous curves) are plotted versus the electron incident energy.}
\label{fig4}  \end{figure}

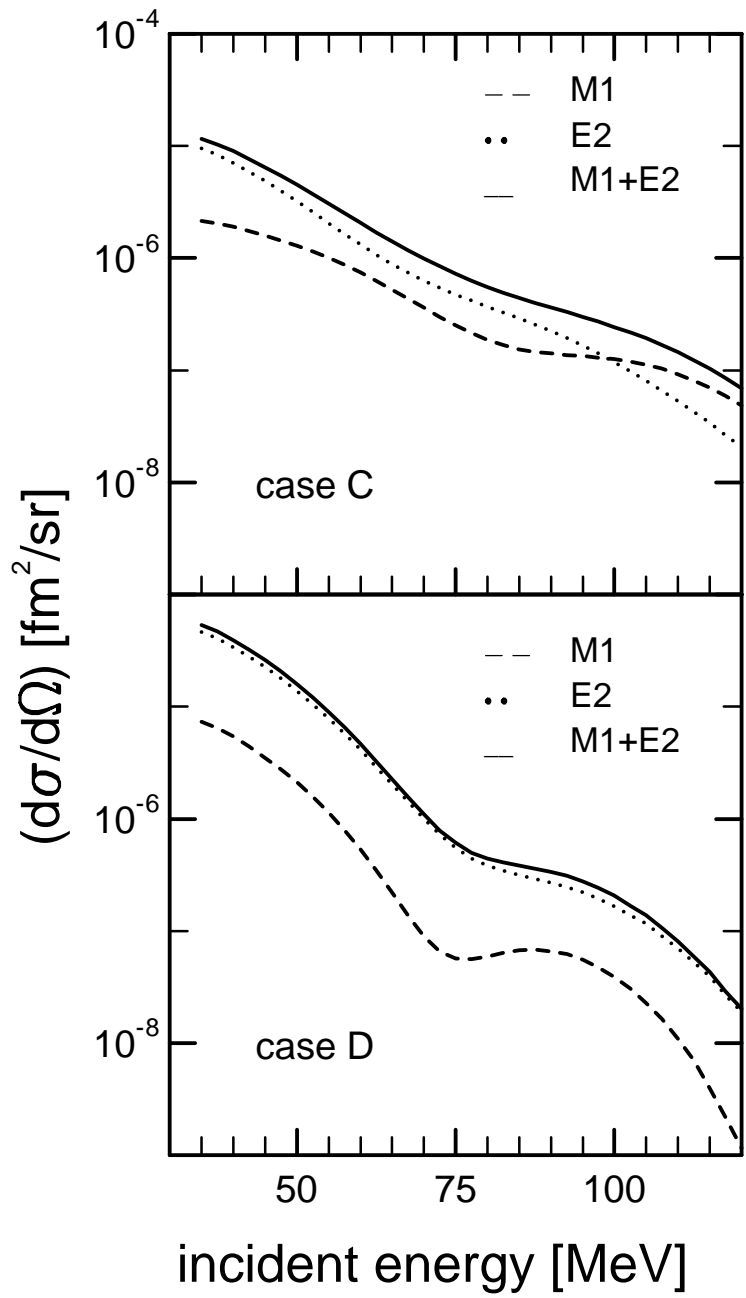
\begin{figure}
\caption{The same as in Fig.\ \protect\ref{fig4}, but for the summed
cross sections of the four strongest M1 excitations around 22 MeV
from the top plot of Fig.\ \protect\ref{fig2} (case C) and the single
state at 22 MeV from the bottom plot of Fig.\ \protect\ref{fig2}
(case D). }
\label{fig5} \end{figure}

   Theoretical $(e,e^{\prime})$ cross sections , calculated in DWBA
\cite{heis} using our RPA wave functions, are plotted in Figs.\
\ref{fig4} and \ref{fig5}.  E2 (dotted curves) and M1 (dashed curves)
contributions to the cross sections for a scattering angle $\theta =
165^{\circ}$ are plotted versus the electron incident energy, together
with their sum (continuous curves).  The $(e,e^{\prime})$ cross
sections of the two strongest high-energy M1 transitions are plotted
in  Fig.\ \ref{fig4}. Case A corresponds to the RPA excitation with
$E_x = 21.55$ MeV and $B(M1) = 0.25 \; \mu^2_N$, while case C refers
to the RPA excitation with $E_x = 20.54$ MeV and $B(M1) = 0.20 \;
\mu^2_N$. Although both states are composed mainly from
configurations with  $\Delta {\cal N}$ = 2, the leading 2q.p.
components in the corresponding RPA wave functions are neutrons with
$\Delta {\cal N}$ = 4 in the former state and protons with $\Delta
{\cal N}$ = 0 in the latter state. Thus, the former excitation has a
more pronounced volume character than the latter one.

   In addition to the above two strongest excitations at high energy
we consider also the two next strongest high-energy orbital M1
excitations from the upper plot of Fig.\ \ref{fig2}.  These four
excitations lie between 20.5 and 23 MeV, have a summed $B(M1)$ value
0.73 $\mu^2_N$, and a summed overlap with the scissors state 11\%.
The sum of their cross sections is displayed in the upper plot (case
C) of Fig.\ \ref{fig5}.  The E2 contribution to the cross section is
dominant over the M1 cross section for incident energies below
100 MeV.  The diffraction structure is smeared out, because the
considered four excitations have very different transition densities.

    The lower plot (case D) in Fig.\ \ref{fig5} corresponds to the
schematic collective excitation at 22 MeV from the lower plot of
Fig.\ \ref{fig2}.  It is seen from Fig.\ \ref{fig5} that even in
backward scattering the cross section is one order of magnitude
larger for E2 than for M1 excitations. The displayed cross sections
are calculated for $\theta = 165^{\circ}$, an angle often used in
experiments, but we have checked that the same E2 dominance is
present even at a fully backward angle.  This result is particularly
interesting in view of the well-known expectation \cite{fagg} for a
strong enhancement of M1 over E2 excitations in backward scattering.
Such a qualitative expectation is based on approximations
(negligible excitation energy compared to the incident electron
energy, PWBA), which do not hold for the DWBA cross section of the
high-energy excitation, dispayed in Fig.\ \ref{fig5}.

   The experimental resolution of single M1 excitations is hindered
furthermore by the strong fragmentation at high energy.  One should
expect, therefore, that above 17 MeV the experimental cross section
of inelastically scattered electrons will originate almost
exclusively from the E2 IVGQR and not from M1 transitions. In
contrast, at low energy the M1 cross section dominates over the E2
cross section \cite{nfd}, at least for low transferred momenta.

\section{Conclusions}

   The high-lying M1 strength is almost purely orbital. It resides in
the same energy region as the isovector E2 strength (IVGQR).
Experimental data on the IVGQR in heavy deformed nuclei are scarce.
The interpretation of the $(e,e^{\prime})$ spectra rely on a rough
estimate of the expected excitation energy of this resonance, based
on the collective model. The main uncertainty arises from the poorly
known isovector potential, which determines also the magnitude of the
isovector coupling in the residual spin-independent interaction, used
in microscopic calculations.  We have adopted here a ratio $r = -2$
between the isovector and isoscalar coupling constants.  The
isoscalar constant is determined microscopically from the condition
of rotational invariance. On the example of $^{160}$Gd, considered
here, the isoscalar coupling gives rise to an isoscalar GQR at 11 MeV
with $B(E2) = 1193 \; e^2fm^4$, in agreement with experiment.

   The adopted isovector ratio, $r = -2$,  produces in RPA an IVGQR
whose $K^{\pi} =1^+$-component is centered at about 22 MeV, but
strongly fragmented over the interval 17--25 MeV. Thus,
even the strongest M1 transition from this region has a relatively
small $B(M1)$ value, 0.25 $\mu^2_N$, in comparison with the strong
low-lying orbital excitations.  The high-lying strengths, summed in
the above interval, are $B(E2) = 1475 \; e^2fm^4, \ \; B(M1) = 3.6 \;
\mu^2_N$. The total RPA energy-weighted E2 strength exhausts exactly
the classical sum rule for $K = \pm 1$.

  The scissors mode is found to fragment over both the low- and
high-energy orbital M1 excitations. Although the latter are more
collective, the coherence of the scissors motion is destroyed at high
energy by the increased fragmentation, due not only to the high level
density, but also to the rotational invariance.  Most of the strongly
fragmented high-lying orbital excitations (17--25 MeV) are not
performing well-pronounced rotational vibrations. This is seen from
their small overlap with synthetic states, where such a geometrical
motion is enforced by construction.  Nevertheless, the orbital
excitations from the above high-energy region have all together a
large overlap, 55\%, with the collective scissors state constructed
from the deformed mean field.  The 1$^+$ excitations below 9 MeV
overlap 77\% with the scissors state. The total scissors overlap of
all the RPA excitations below 30 MeV is 1.46.  The excess above unity
is produced by the isovector coupling, corresponding to a stronger
attractive neutron-proton interaction, which enhances the scissors
motion.

   The recent canonical quantization of the isovector rotor in
relative conjugate variables \cite{irm} provides simple analytical
results in the deformed oscillator basis. They strongly support the
interpretation of all the orbital M1 strength, at both low and high
energies, as manifestation of the fragmented collective scissors
mode.

   The $(e,e^{\prime})$ cross sections of 1$^+$ states above 17
MeV originate mainly from E2 excitations even for backward
scattering, while the contributions of the high-lying orbital M1
excitations to these cross sections are typically one order of
magnitude smaller.

   Thanks are due to Jochen Heisenberg for providing us with his DWBA
code. This work is supported by the Deutsche Forschungsgemeinschaft
(DFG).

\newpage
\begin{table}
\caption{ Comparison of different theoretical QRPA predictions on
the summed high-energy M1($\uparrow$) strength and overlap with the
scissors state {\bf R}$^{\dag}\vert \rangle$ \protect\cite{nfscis}
in rare-earth nuclei. }

\begin{tabular}{lcccl}

Nucleus &  E    & $\sum$ B(M1) & $\sum$ EB(M1) &
$\sum \vert \langle{\rm \bf R}1^+\rangle \vert ^2$  \\
   & [MeV] & [$\mu^2_N$] & [MeV$\; \mu^2_N$] & \quad [\%]  \\
\tableline
$^{160}$Gd $^a$ & 17--25 & 3.6 & 76 & \quad 55 $^b$  \\
$^{164}$Dy $^c$ & 21--23 & 3.6 & 79 & \quad 30       \\
$^{152}$Dy $^d$ &   27   & 1.5 & 40 & \quad 40       \\
\end{tabular}
\end{table}
\noindent $^a$ This work. \\
\noindent $^b$ 38\% of the total overlap 1.46. \\
\noindent $^c$ Ref. \cite{zaw90}. \\
\noindent $^d$ Ref. \cite{hana}. \\

 \newpage
\narrowtext

\end{document}